\documentclass[runningheads,orivec]{llncs}
\usepackage[T1]{fontenc}
\usepackage{multirow}
\usepackage{tablefootnote}
\usepackage{subfig}
\usepackage{svg}
\usepackage{subcaption}
\usepackage{graphicx}
\usepackage{amsmath}
\usepackage{orcidlink}

\usepackage{cleveref}
\begin{document}
\title{Designing Effective LLM-Assisted Interfaces for Curriculum Development}
\titlerunning{Designing Effective LLM-Assisted Interfaces for Curriculum Development}

\author{
  Abdolali Faraji \inst{1} \orcidlink{0000-0002-3557-9345} \and
  Mohammadreza Tavakoli \inst{1} \orcidlink{0000-0002-7368-0794} \and
  Mohammad Moein \inst{1} \orcidlink{0000-0002-3285-8226} \and
  Mohammadreza Molavi \inst{1} \orcidlink{0009-0006-0423-0729} \and
  Gábor Kismihók \inst{1} \orcidlink{0000-0003-3758-5455}
}

\authorrunning{A. Faraji et al.}

\institute{Leibniz Information Centre for Science and Technology (TIB)
    \\\email{\{abdolali.faraji, reza.tavakoli, mohammad.moein, mohammadreza.molavi, gabor.kismihok\}@tib.eu}}
\maketitle              % typeset the header of the contribution
\begin{abstract}

\textit{Large Language Models} (\textit{LLMs}) have the potential to transform the way a dynamic curriculum can be delivered. However, educators face significant challenges in interacting with these models, particularly due to complex prompt engineering and usability issues, which increase workload. Additionally, inaccuracies in LLM outputs can raise issues around output quality and ethical concerns in educational content delivery. Addressing these issues requires careful oversight, best achieved through cooperation between human and AI approaches. This paper introduces two novel User Interface (\textit{UI}) designs, \textit{UI Predefined} and \textit{UI Open}, both grounded in \textit{Direct Manipulation} (\textit{DM}) principles to address these challenges. By reducing the reliance on intricate prompt engineering, these UIs improve usability, streamline interaction, and lower workload, providing a more effective pathway for educators to engage with LLMs. In a controlled user study with 20 participants, the proposed UIs were evaluated against the standard ChatGPT interface in terms of usability and cognitive load. Results showed that \textit{UI Predefined} significantly outperformed both \textit{ChatGPT} and \textit{UI Open}, demonstrating superior usability and reduced task load, while \textit{UI Open} offered more flexibility at the cost of a steeper learning curve.  These findings underscore the importance of user-centered design in adopting AI-driven tools and lay the foundation for more intuitive and efficient educator-LLM interactions in online learning environments.

\end{abstract}
\section{Introduction}
Online learning has become indispensable in modern education, revolutionizing the way knowledge is accessed and disseminated. Its popularity has surged in recent years due to several factors. The convenience of learning at one's own pace and from any location has made it particularly appealing to busy individuals and those with limited access to traditional educational institutions \cite{gros2023future}.
Moreover, the demand for rapid access to up-to-date information on various topics, including skills, tasks at hand, and hobbies, has made online learning a highly effective and efficient educational method \cite{tavakoli2022ai}. This highlights the importance of dynamic up-to-date curriculum development in the context of online learning \cite{tavakoli2022hybrid}. Maintaining curricula dynamically to align with the rapid pace of societal changes and learning needs poses significant challenges \cite{tavakoli2022hybrid}. The burden on teachers to continuously update their content is substantial, as it requires extensive time and effort \cite{denny2023can}. Additionally, accessing high-quality, valid information from the vast amount of available online resources is also challenging, making it difficult to ensure the accuracy and relevance of the curriculum content \cite{abdi2021evaluating}.

The emergence of LLMs has sparked optimism for addressing the challenges associated with dynamic curriculum development \cite{moein2024beyond}. Despite the potential of LLM-based methods, they are questionable in terms of the output quality \cite{denny2023can,bahrami2024llm} (i.e. inaccurate, wrong, or even ethically questionable outputs). This has led to a growing interest in investigating the collaborative Human-AI approaches that combine the strengths of LLMs with human expertise \cite{tavakoli2022hybrid}. Such collaboration between AI and human educators enables a sustainable, future-ready approach to curriculum development \cite{padovano2024towards}.

One of the primary limitations of LLM usage by experts lies in their text-based UIs, which are often not user-friendly and impose a higher cognitive load \cite{feng2024canvil}. Interacting with these text-based UIs relies heavily on textual prompts that need to be optimized. This technology-centered method (in contrast to the user-centered method) requires \emph{careful wording} and \emph{precise references} to the object of interest, which results in more difficult optimization \cite{chen2023unleashing}. This leads to challenges like 1) \emph{indirect engagement} due to limited direct access to the object of interest, 2) \emph{semantic distance} in expressing intent in written form, and 3) \emph{articulatory distance} between prompts and their intended actions \cite{masson2024directgpt}. Addressing these challenges through user-friendly LLM interfaces is essential to facilitate content development and maximize AI's potential \cite{denny2023can,masson2024directgpt}.

This paper proposes two novel UI designs to address the challenges of LLM-based curriculum development, specifically those related to their interfaces. These proposed UIs aimed to optimize the interaction between educators and LLMs, potentially surpassing the capabilities of existing interfaces. Furthermore, by conducting a comprehensive evaluation, we sought to determine whether these UI solutions could enhance the efficiency and effectiveness of curriculum development processes in terms of usability and workload. Therefore, our main contributions are:
\begin{itemize}
    \item Designing and prototyping UIs for LLM, specifically ChatGPT, which aim at enhancing the usability of LLMs for educators, when it comes to curriculum development
    \item Conducting an experiment to evaluate the proposed UIs against the default ChatGPT UI in terms of usability, effectiveness, and cognitive load
\end{itemize}

\section{Related Work} \label{sec:state}
In this section, we cover the related research, which influenced key components of our conceptual and technical work. We begin by discussing AI-based curricula development. Next, we explore the development of LLM-based UIs. Finally, we examine the research on human-computer interaction within the context of AI systems, focusing on how to optimize the user experience.

\subsection{Curricula Development}
Previous research has explored the integration of AI algorithms into curriculum development. For instance, \cite{Mariani_Pellegrini_De_Momi_2021} used an adaptive AI algorithm to create real-time curricula for robot-assisted surgery using simulators, based on learners' feedback. Although the number of participants was limited, their results demonstrated improved learning outcomes for students using these AI-generated curricula. Additionally, both \cite{molavi2020extracting} and \cite{Kawamata_Matsuda_Sekiya_Yamaguchi_2021} employed \textit{Latent Dirichlet Allocation (LDA)} to extract covered topics from educational resources to support curriculum development processes. Moreover, \cite{tavakoli2022hybrid} utilized various traditional machine learning models (e.g., LDA and Random Forest) to assist human experts in developing curricula by recommending learning topics and related educational resources. It can be observed that previous research efforts primarily relied on traditional machine learning methods, limiting their focus to specific educational domains.

\subsection{LLM-based UIs}
To address the challenges of interacting with LLMs, researchers have explored various UI approaches. For instance, \cite{masson2024directgpt} proposed the \textit{DirectGPT} UI to enhance the usability of LLMs by providing features like continuous output representation and prompt control. Through a user study with 12 experts, they demonstrated that \textit{DirectGPT} users achieved better task completion times and reported higher satisfaction compared to traditional interfaces. Additionally, \cite{dang2022ganslider} explored the impact of different slider types on user control over generative models, highlighting the importance of UI design for effective interaction.

Beyond general-purpose LLMs, researchers have also investigated LLM-based UI applications in specific domains. \cite{liu2024sprout} developed \textit{SPROUT}, a tool designed to assist users in creating code tutorials using LLMs. Their study found that SPROUT significantly improved user satisfaction and the quality of generated content. Furthermore, \cite{wang2024lave} demonstrated the potential of LLMs in video editing workflows, emphasizing the need for intuitive UIs to facilitate this process.

\subsection{Human-Computer Interaction for AI Systems}
%%%  and now the same problem has happened
While AI solutions, particularly LLMs, have opened up numerous possibilities that were previously unimaginable, the development of many AI-based systems continues to be hindered by a "technology-centric" approach rather than a "user-centric" approach \cite{xu2023transitioning,ozmen2023six}.

As already mentioned in the introduction, interaction with LLMs usually happens using textual prompts which introduce different challenges (i.e. indirect engagement, semantic distance, and articulatory distance). These challenges closely resemble issues that caused the shift from command-line interfaces to \textit{Direct Manipulation (DM)} interfaces (like graphical user interfaces) in the 1980s \cite{hutchins1985direct}. Shneiderman \cite{shneiderman1983direct} defined DM interfaces with four key traits: 
\begin{enumerate}
    \item[(DM1).] Continuous representation of the object of interest
    \item[(DM2).] Physical actions (e.g. movement and selection by mouse, joystick, touch screen, etc.) instead of complex syntax
    \item[(DM3).] Rapid, incremental, reversible operations whose impact on the object of interest is immediately visible
    \item[(DM4).] Users should recognize the actions they can do instead of learning complex syntax
\end{enumerate}

We use the term \textit{DM} in our paper to refer to these four pillars in the user interface.

\subsection{Lessons Learned}
As discussed in the Curricula Development subsection, there is a clear need for developing and evaluating more advanced and up-to-date AI-driven solutions (such as those based on large language models) for curriculum development, given their demonstrated potential in these educational contexts \cite{denny2023can}. However, as we highlighted, these solutions must be complemented with user-friendly interfaces, as text-based interaction with LLMs can be less than ideal for users \cite{masson2024directgpt}. Consequently, UI principles (e.g., DM), as explored earlier, should be carefully considered when designing such interfaces for LLM-based curricula development.

\section{Method} \label{sec:method}

To address text-based interface challenges, we applied DM principles to UIs designed for \emph{course outline creation}, assisting educators in defining course titles, learning outcomes, and related topic lists. Below, we describe the two developed UIs, the prompt engineering techniques used, and the control group's standard ChatGPT UI.

\subsection{UI 1: Predefined Commands}

\begin{figure}[h!]
    \centering
    \includegraphics[width=\linewidth]{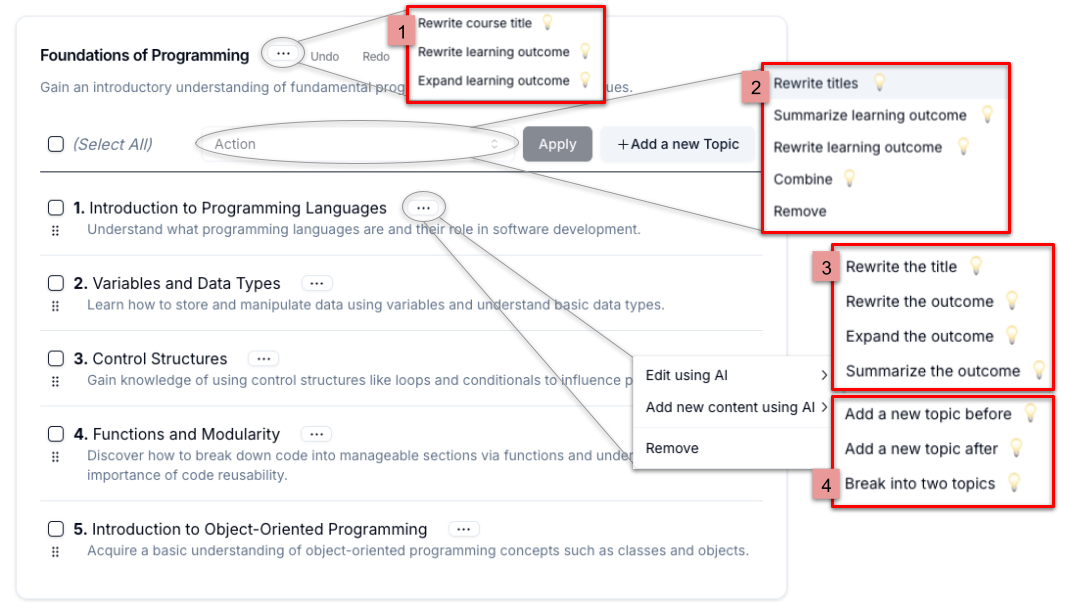}
    \caption{An example of a course outline representation in \emph{UI Predefined}. Four groups of predefined commands are visible: \textit{group-1)} course-related commands, \textit{group-2)} batch commands, \textit{group-3)} single-topic refinement commands, and \textit{group-4)} content generation with respect to a single topic.} 
    \label{fig:ui1outline}
\end{figure}

\emph{UI Predefined} offers educators clickable buttons with predefined commands in a familiar graphical user interface (\textit{GUI}) layout. These commands, curated from expert inputs in European Union educational projects\footnote{Project details can be found here: \url{https://www.tib.eu/en/research-development/research-groups-and-labs/learning-and-skill-analytics/projects}}, support key tasks in creating and refining course outlines. Commands in \textit{UI Predefined} are organized into four functional groups, presented as clickable actions, as depicted in \Cref{fig:ui1outline}. The first group includes commands for modifying course-level information (i.e. the course title and learning outcome). The second group comprises batch actions, which are activated upon selecting multiple topics. The third group focuses on actions for refining individual topics, while the fourth group facilitates the generation of new content based on a single topic.

As it is visible in \Cref{fig:ui1outline}, the course outline representation in \emph{UI Predefined} is enhanced with an interactive table format (aligned with DM1) which replaces static text, improving clarity and interaction. Command execution is transparent, with only the updated outline sections displaying a loading effect. A meatball menu next to each item supports contextual actions (DM2, DM4), while checkboxes enable batch operations (DM3). Users can directly edit course titles or outcomes, rearrange topics with drag-and-drop, and add/remove topics, aligning with DM2 and DM4. Undo/redo functionality supports rapid, reversible interactions consistent with DM3.

\subsection{UI 2: Open Commands}
\begin{figure}[h]
    \centering
    \includegraphics[width=\linewidth]{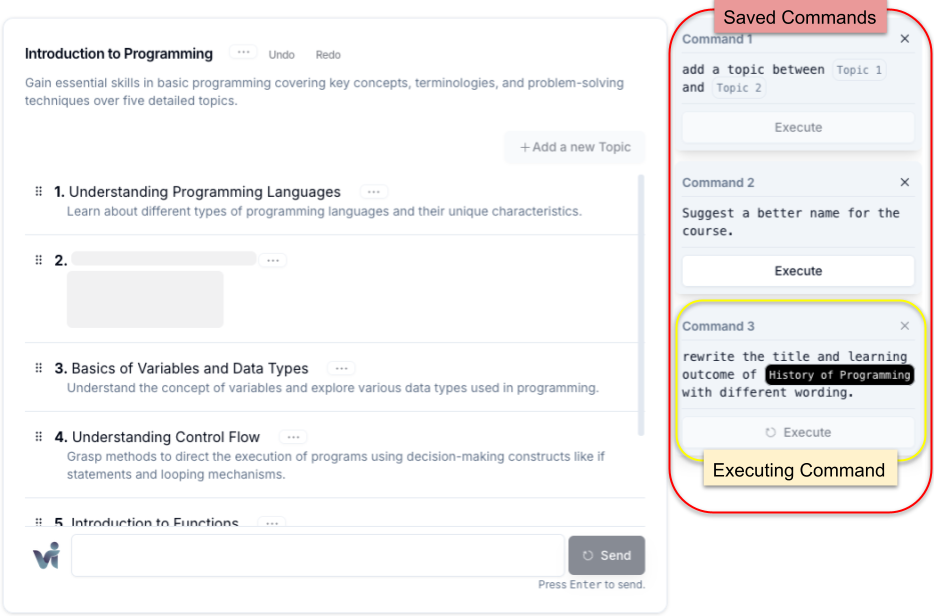}
    \caption{An example of a course outline being developed by a user with dynamic commands in \emph{UI Open}. In this figure, Command 3 is being executed on Topic 2.}
    \label{fig:ui2-complete}
\end{figure}

\emph{UI Open} was developed as an alternative to \emph{UI Predefined}, retaining its core visual structure (as shown in Figure \ref{fig:ui2-complete}) while addressing key limitations, particularly the lack of direct user interaction with the LLM. To overcome this limitation, a dynamic command mechanism was introduced, allowing users to engage directly with the LLM. These command functions include three key enhancements: (1) automatic integration of the course outline context, ensuring LLM awareness of the structure; (2) a drag-and-drop interface for associating commands with specific course elements; and (3) the ability to save and reuse commands, reducing repetition and improving time-efficiency.

\begin{figure}[h]
    \centering
    \includegraphics[width=\linewidth]{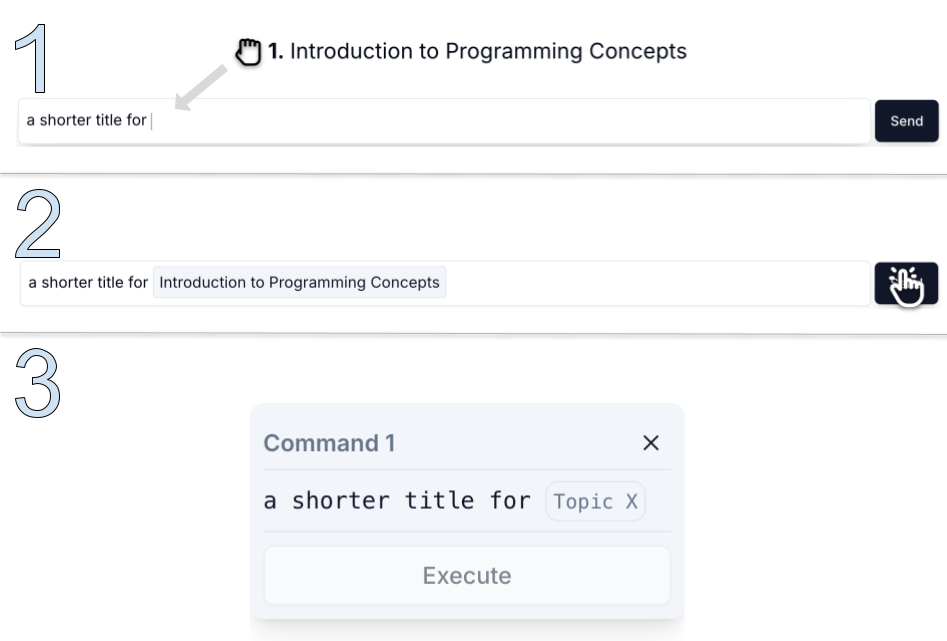}
    \caption{\emph{UI Open}: \emph{1)} User is creating a command while dragging a topic for localization \emph{2)} Command is executed and saved \emph{3)} Any other topic can be dragged to the saved command.}
    \label{fig:dynamic-commands}
\end{figure}

To enhance user-friendliness, \emph{UI Open} incorporates a chat box at the bottom of the course outline interface. The outline representation remains largely unchanged (DM1), allowing manual edits via a click-based interface (DM2). Unlike \emph{UI Predefined}, it removes predefined buttons, offering a more flexible approach. The chat box enables users to input, execute, and save dynamic commands for reuse, supporting DM2 and DM3. Commands can be global (affecting the entire outline) or local (applied to specific sections), adjusted by dragging topics into the chat box (DM2), as shown in Figure \ref{fig:dynamic-commands}.

\subsection{Prompt Engineering}
To achieve the desired results from an LLM in our UIs, prompts were tailored based on OpenAI's best practices for prompt engineering \cite{openaiprompt}. To be more specific, using the \texttt{gpt-4o-2024-08-06} model (the latest model at the time of the experiment), a system prompt was applied to (1) define the LLM's persona, (2) delimit prompt components, (3) specify the task (course outline curation), and (4) establish the output format. Moreover, the current course outline and user-issued commands were always included in prompts, allowing the LLM to respond within context. The output format adhered to a defined structure, translated by the UI engine for display. To ensure consistency, each topic generated by the LLM was assigned an identifier, enabling precise referencing throughout interactions.

\subsection{ChatGPT Replica Tool}
To establish a control application for comparison with \emph{Predefined} and \emph{Open} UIs, we required a tool that emulated the ChatGPT interface. Direct usage of ChatGPT was unsuitable, as it did not allow for monitoring user interactions, nor did it guarantee the use of the same model utilized in our UIs. After exploring alternatives, we opted for \texttt{open-webui}\footnote{open-webui repository: \url{https://github.com/open-webui/open-webui}}, an open-source project designed for interacting with LLMs. This tool offers an experience closely comparable to the original ChatGPT interface, making it a suitable choice for our experiment.

\section{Evaluation} \label{sec:evaluation}
We conducted a user study to evaluate the effectiveness and usability of the UIs: \textit{UI Predefined}, \textit{UI Open}, and a control interface replicating \textit{ChatGPT}\footnote{A anonymized demo of the experiment, including different steps, is available under: \url{https://figshare.com/s/faa7fcea8bbf9d93ad1f}}. Participants from diverse educational backgrounds used each UI and completed a questionnaire capturing their experience and background. Below, we detail the experimental design, procedures, and analytical methods.

\subsection{Participants}
\begin{figure}[h]
    \centering
    \subfloat[\centering Age]{{\includegraphics[width=0.5\textwidth]{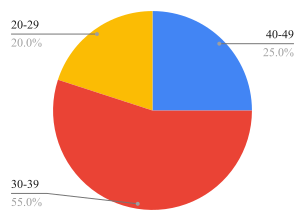}}}
    \subfloat[\centering Gender]{{\includegraphics[width=0.5\textwidth]{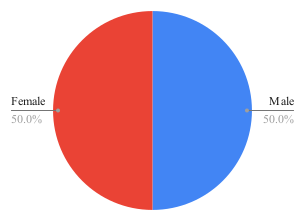}}}
    \\
    \subfloat[\centering LLM Experience - {\scriptsize self rated from 1 (least) to 5 (most)}]{{\includegraphics[width=0.5\textwidth]{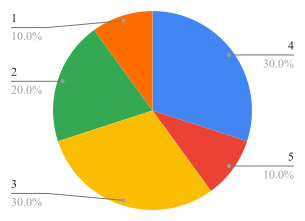}}}
    
    \caption{Background information of the participants.}
    \label{fig:demography}
\end{figure}

The study involved 20 participants, with teaching experience ranging from 1 to 21 years, providing a broad spectrum of perspectives on the usability and effectiveness of the UIs. They were evenly distributed by gender, with an equal number of females and males. Each participant had a minimum of five years of professional experience across diverse fields, including but not limited to Computer Science, Mathematics, Nursing, and Soft Skills. Additionally, participants exhibited varying degrees of familiarity with \textit{ChatGPT}, ranging from low to very high proficiency. This range of teaching and technological backgrounds enabled us to collect feedback that reflects a diverse set of educational contexts. The demographic distribution of the participants is depicted in Figure \ref{fig:demography}.

\subsection{Questionnaire}
The evaluation focused on two primary objectives: (1) assessing perceived workload and (2) measuring interface usability. We used the \textit{Raw NASA Task Load Index} (NASA RTLX) to quantify task load and the \textit{System Usability Scale} (SUS) to evaluate usability, both established methods in HCI research \cite{kosch2023survey,bangor2008empirical}. \textit{NASA RTLX} measures perceived workload in the dimensions of mental demand, physical demand, temporal demand, performance, effort, and frustration. \textit{SUS} on the other hand, covers the ease of use, efficiency, consistency, and learnability. Data was collected through a three-step questionnaire: participants first completed the NASA RTLX to assess workload, followed by the SUS for usability evaluation, and finally answered an open-ended question for additional feedback. This approach combined quantitative and qualitative insights into the UIs' effectiveness and usability.

\subsection{Experiment Steps}
Each participant followed a structured process to evaluate the three AI-powered UIs for course outline development:  
\begin{enumerate}
    \item \textbf{Study Introduction ($\approx$ 10 min).} Participants were briefed on the study’s objectives, focusing on the workload and usability of the UIs.  
    \item \textbf{Background Information ($\approx$ 5 min).} Demographic data, including teaching experience and prior \textit{ChatGPT} familiarity, were collected.  
    \item \textbf{Reference Course Outline ($\approx$ 30 min).} Participants created a baseline course outline in their teaching subject for comparison with different UI outputs.  
    \item \textbf{Curriculum Development with UIs (3 rounds, $\approx$ 30 min each).} Participants used each of the three UIs in randomized order, for a fair evaluation, spending 20 minutes per UI to create course outlines, followed by completing an evaluation form.  
    \item \textbf{Final Evaluation ($\approx$ 5 min).} Participants provided suggestions for improvement and shared challenges.
\end{enumerate}

\section{Results and Discussion} \label{sec:result}
\subsection{Results}
All 20 participants completed the study\footnote{The raw data, including the background information of the participants, the questionnaires, and the participants' answers, is available at: \url{https://figshare.com/s/76c780adbae8f3d22f07}}. Results showed \emph{UI Predefined} significantly outperformed \emph{ChatGPT} in workload, usability, and efficiency, while \emph{UI Open} ranked second with a non-significant improvement over \emph{ChatGPT}. We used the \textit{Wilcoxon signed-rank test} \cite{woolson2005wilcoxon} to analyze differences in workload and usability, with statistical analysis conducted in \textit{Python} 3.12 using \textit{scipy}. \textit{p-values} were reported to assess significance.

\subsubsection{Workload and Performance}
\begin{figure}[h]
    \centering
    \includegraphics[width=\linewidth]{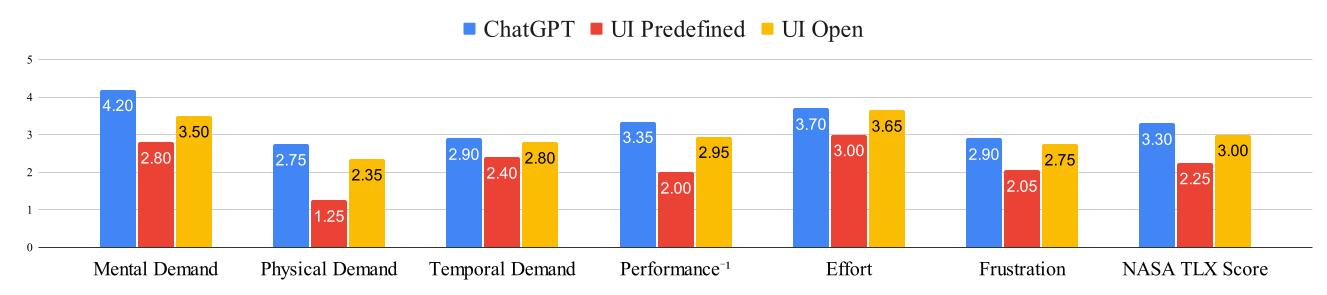}
    \caption{The average results for each metric in NASA RTLX. Lower values mean a lower workload. {\scriptsize ($\text{Performance}^{-1}$ is defined as $8 - \text{Performance}$)}}
    \label{fig:nasa-tlx}
\end{figure}

To compute the NASA RTLX scores, we transformed the \textit{Performance} dimension into \textit{Performance$^{-1}$} by subtracting the original value from 8, ensuring lower scores indicated better performance (Performance$^{-1}$ would be between 1 to 7 like other dimensions). \textit{UI Predefined} had the lowest workload with a mean score of 2.25, followed by \textit{UI Open} (3.00) and \textit{ChatGPT} with the highest workload (3.30). The adjusted \textit{Performance} metric further emphasized the efficiency gap, with \textit{ChatGPT} showing a consistently higher workload compared to both developed UIs. These findings highlight the superior efficiency and reduced cognitive load of \textit{UI Predefined} and \textit{UI Open} over the control. Detailed results and \textit{p-values} are presented in Table \ref{tab:nasa-results} and visualized in Figure \ref{fig:nasa-tlx}.

\begin{table}[h]
    \centering
    \begin{tabular}{c c|c|c}
         \multicolumn{2}{c|}{Comparison} & $\lvert\text{Difference}\rvert$ & \textit{p-value} \\
         \hline
         
         \textit{\textbf{UI Predefined}} & ChatGPT  & \multirow{2}{*}{1.05} & \multirow{2}{*}{< 0.03} \\
         \textbf{2.25} & 3.30 & & \\ \hline

          \textit{\textbf{UI Predefined}} & \textit{UI Open}  & \multirow{2}{*}{0.75} & \multirow{2}{*}{< 0.02} \\
         \textbf{2.25} & 3.00 & & \\ \hline

          \textit{\textbf{UI Open}} & \textit{ChatGPT}  & \multirow{2}{*}{0.30} & \multirow{2}{*}{-} \\
         \textbf{3.00} & 3.30 & & \\ \\
    \end{tabular}
    \caption{Workload Comparison.}
    \label{tab:nasa-results}
\end{table}

\subsubsection{Usability}
\begin{figure}[h]
    \centering
    \includegraphics[width=\linewidth]{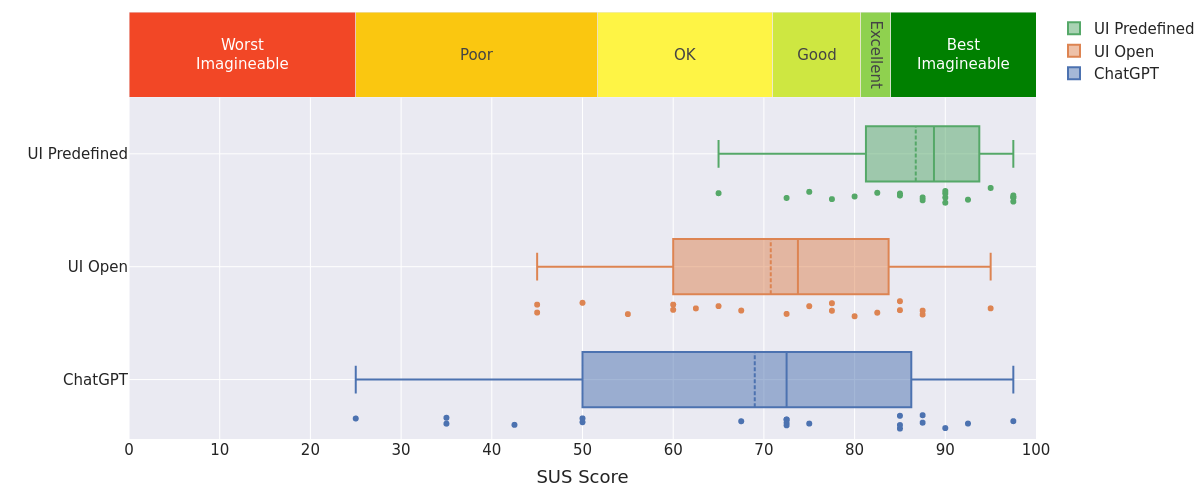}
    \caption{UIs' SUS Scores. Min, max, 1st, 2nd, and 3rd quartiles are shown. The dotted lines are $mean$.}
    \label{fig:sus}
\end{figure}

The System Usability Scale (SUS) results show \textit{UI Predefined} achieving the highest usability score of 86.75, well above the "Excellent" threshold \cite{bangor2008empirical}. In comparison, \textit{UI Open} scored 70.75, and \textit{ChatGPT} scored 69.00, both within the "OK" category. These findings highlight the superior ease of use and participant satisfaction with \textit{UI Predefined}. Detailed \textit{p-values} are in Table \ref{tab:sus-results}, with results visualized in Figure \ref{fig:sus}.

\begin{table}[h]
    \centering
    \begin{tabular}{c c|c|c}
         \multicolumn{2}{c|}{Usability Comparison} & $\lvert\text{Difference}\rvert$ & \textit{p-value} \\
         \hline

        \textit{\textbf{UI Predefined}} & ChatGPT  & \multirow{2}{*}{17.75} & \multirow{2}{*}{< 0.009} \\
         \textbf{86.75} & 69.00 & & \\ \hline

         \textit{\textbf{UI Predefined}} & UI Open  & \multirow{2}{*}{16.00} & \multirow{2}{*}{< 0.002} \\
         \textbf{86.75} & 70.75 & & \\ \hline

          \textit{\textbf{UI Open}} & ChatGPT  & \multirow{2}{*}{1.75} & \multirow{2}{*}{-} \\
         \textbf{70.75} & 69.00 & & \\ \\
    \end{tabular}

    \caption{SUS Comparison.}
    \label{tab:sus-results}
\end{table}

\subsection{Discussion}
Our findings indicated that the UI with predefined commands outperformed the others in both usability (SUS) and workload (NASA RTLX) dimensions. This superiority was attributed to its ability to streamline the course outline creation process. Eight participants specifically noted that predefined commands significantly reduced their typing burden, while ten emphasized the overall ease of use. 

While the UI with open commands also received positive feedback for its flexibility, participants acknowledged a steeper learning curve. This suggests that a hybrid approach, combining predefined and open commands, might offer the optimal balance of usability and adaptability. Although the ChatGPT UI received some preference due to participants’ familiarity, all participants agreed that features like command definition, reuse, and better output presentation were essential for enhancing their overall experience.

Despite the promising results of our study, there are limitations that must be acknowledged. The experiment was constrained to controlled environments, which may not accurately reflect the dynamic and diverse settings in which educators interact with LLMs in real-world scenarios. On the other hand, the study was conducted on a limited number of educators, potentially limiting its diversity. However, we tried to mitigate this limitation by inviting educators from diverse areas, ensuring representation across different backgrounds and demographics. These limitations suggest that future research should encompass a wider range of user backgrounds and real-world testing to enhance the applicability and robustness of UI designs in practical educational contexts.

\section{Conclusion} \label{sec:conclusion}

In conclusion, this paper highlighted the growing role of online learning and the potential of LLMs in education. While LLMs enhance accessibility and personalization, they also pose challenges, particularly for educators. Key concerns include (1) the accuracy of LLM outputs and (2) the complexity of effective interaction. To address these, we proposed and evaluated two user interfaces, \textit{UI Predefined} and \textit{UI Open}, designed with DM principles to simplify educator interaction. These interfaces reduce reliance on prompt engineering and improve usability. Additionally, they mitigate the risks associated with output quality by fostering Human and AI collaboration, integrating a human check to ensure accuracy and alignment with the teacher standards. 

Our findings revealed that the UI Predefined significantly outperformed ChatGPT, showcasing its potential to streamline educational interactions with a set of expert-derived commands that reduce cognitive load and enhance efficiency and usability. Although UI Open also showed improvements over ChatGPT by allowing more direct, context-rich interactions for educators, it did not exhibit a statistically significant advantage.

Building on our findings, future research should explore a hybrid approach that integrates elements of both UI Open and UI Predefined, balancing structured guidance with open-ended interaction. Additionally, expanding the study to a more diverse group of educators across various real-world settings will enhance the generalizability of the results. Finally, the impact of different UI designs on the learning processes should be investigated. This will help to improve learning as the ultimate goal of educational applications.

\bibliographystyle{splncs04}
\bibliography{references}
\end{document}